\documentclass[a4paper,12pt]{article}

\newcommand{\be}{\begin{equation}}
\newcommand{\ee}{\end{equation}}
\newcommand{\bea}{\begin{eqnarray}}
\newcommand{\eea}{\end{eqnarray}}

\usepackage[dvips]{graphicx}

\begin{document}

{\Large\bf\noindent Non-Spherical  Newtonian Gravitational Collapse:\\
a Set of Exact Nonlinear Closed--Form Solutions}

\vspace{0.5cm}

{\centerline{A.~D.~Chernin$^{1}$, A.~S.~Silbergleit$^{2}$}}

\vspace{2mm}

{\it $^1$Sternberg Astronomical Institute, Moscow University,
Moscow, Russia}. 

e-mail: {chernin@sai.msu.ru}

{\it $^2$HEPL, Stanford University, Stanford, CA 94305-4085, USA}. 

e-mail: alex.gleit@gmail.com

\vspace{10mm}

 \noindent

\section*{Abstract}

We give a set of exact nonlinear closed--form solutions for the
non-spherical collapse of pressure-less matter in Newtonian
gravity, and indicate their possible cosmological applications.

{\it Keywords:} Newtonian gravitation: free collapse, large-scale
cosmic structures

\section{Introduction\label{s1}}

The Zeldovich approximation~\cite{A0}, of  decades ago, serves as
a reliable basis for modern studies of the large-scale  cosmic
structure formation  (see the classic paper~\cite{A1}, recent works~\cite{A2},
and the references  therein).  It stemmed 
from an elegant  exact nonlinear solution  for  the Newtonian anisotropic motion 
of dust-like gravitating matter, which describes a growth of initially weak adiabatic perturbations due to 
gravitational instability in an expanding universe. 

We re-visit  the Zeldovich solution and its generalizations~\cite{2, 3} to reveal their new features emerging under the time-reversal 
transformation (TrT), $t \longrightarrow - t$. To start with, we note that  Friedmann's general--relativistic cosmological solution is known to have a Newtonian analog 
corresponding to a uniform isotropic expansion of a perfect pressure-less gravitating fluid (`dust'). After the TrT, it turns to the solution describing the isotropic gravitational collapse (parabolic motion):
\be\label{isotr}  
\vec x(t,\vec\xi) = (t_0-t)^{2/3}\vec\xi; \qquad \rho (t) = [6\pi G(t_0-t)^2]^{-1}\; .
\ee
Here $\vec x$ is the Euler, and $\vec\xi$ is the
Lagrange coordinate, $G$ is the Newtonian gravitation constant,
$\rho(t)$ is the matter density, and $t_0$ is the collapse time, when the scale factor $a(t) =
(t_0-t)^{2/3}$ becomes  zero and the density goes to infinity. Solution~(\ref{isotr}) applies to uniform spheres of a finite total mass and any initial radius.

In this paper we give a set of solutions for the non-spherical (anisotropic) collapse 
using the time reversal of expansion solutions found in~\cite{A0,2,3}, and discuss their properties. The main reason for this is that
in modern cosmology three types of building blocks are recognized 
forming the large-scale Cosmic Web. Those are rich spherical
clusters of galaxies, flat superclusters (`pan-cakes'), and
elongated filaments. With all the necessary reservations, the formation and evolution of spherical clusters can hopefully be described by the asymptotically spherical collapse solution~(\ref{decx}). Superclusters and filaments are seen as highly oblate and highly prolate
structures, respectively. Their evolution may be
described by the solutions ~(\ref{grx}),(\ref{grr}) and~(\ref{genx}),(\ref{genr}) containing the growing mode, assuming that they correctly expose the basic features of a real finite mass collapse with the same dynamical asymptotics. Generalized formulas ~(\ref{Lam})  add dark energy to the picture.

One can construct the collapse solutions  directly, without the TrT; this  explains the powers of  $\tau=t_0-t$ involved in them. Assume that the flow ~(\ref{isotr}) is perturbed in one direction by a term with separated variables: 
\be\label{motlawgen}
x_1=(\tau)^{2/3}\xi_1+g(\tau)f(\xi_1);\qquad x_{2,3}=(\tau)^{2/3}\xi_{2,3}\; ;
\ee
functions $g(\tau)$ and $f(\xi_1)$ are so far arbitrary. By repeatedly differentiating these expressions one obtains the accelerations  $\partial^2x_n/\partial t^2,\; n=1,2,3$; according to the equations of motion, they are the negative of the respective components of the gravitational potential gradient. The divergence of the gradient produces the Laplacian of the potential; divided by $4\pi G$, it gives, by the Newton law of gravity, the first, `gravitational', formula for the matter density. The other one comes from the mass conservation, i.e., the continuity equation;  the two densities coincide when 
\be\label{Eqg}
g^{''}-(4/9\tau^2)g=0\; ;
\ee
linearly independent solutions of this equation  are $\tau^{4/3}$ and $\tau^{-1/3}$. The first one corresponds to the solution~(\ref{decx})  of sec.~\ref{s2}, the second gives the solution~(\ref{grx}), sec.~\ref{s3}. A linear combination, $\tau^{4/3}f(\xi_1)+\tau^{-1/3}F(\xi_1)$, turns out also possible in the expression~(\ref{motlawgen}) for $x_1$, allowing for the solution~(\ref{genx}) of sec.~\ref{s4}.

\section{Collapse with decaying asphericity\label{s2}}

The well--known Zeldovich's `pan-cake' solution~\cite{A0} for planar motion of
pressure-free matter takes the form, after the TrT :
\bea \label{decx}
x_1(t,\xi_1) = (t_0-t)^{2/3}\xi_1 + (t_0-t)^{4/3} f(\xi_1),\qquad x_{2,3}(t,\xi_{2,3})=(t_0-t)^{2/3}\xi_{2,3}\; ;\\
\label{decr}
\rho(t,\xi_1) = \frac{1}{6\pi G (t_0-t)^2} \,\frac{1}{1 + (t_0-t)^{2/3}f^{'}(\xi_1)},\qquad f^{'}(\xi_1)=\frac{df}{d\xi_1}\; .\quad
\eea
The  deviation in $x_1$ from the spherically symmetric motion~(\ref{isotr}) goes to zero when $t\to t_0-0$, becoming negligible against the background (a decaying mode). So formulas~(\ref{decx}),(\ref{decr}) describe an initially non-spherical collapse which undergoes spherization  and becomes completely spherical in the end.

The density~(\ref{decr}) is non-singular until the collapse when $1 + (t_0-t)^{2/3}f^{'}(\xi_1)>0$ for all $\xi_1$,  guaranteeing also a unique inverse, $\vec \xi=\vec \xi(t,\vec x)$, of the law of motion $\vec x=\vec x(t,\vec\xi) $,  as required. The inequality, and thus the solution, is apparently valid for a {\it finite} period of time, $t_i<t<t_0$, if $f^{'}(\xi_1) \geq -(t_0-t_i)^{-2/3}$. For $t_i=-\infty$, function $f(\xi_1)$ is non--decreasing, $f^{'}(\xi_1) \geq0$, so the solution holds on the whole semi--axis  $-\infty<t<t_0$.

\section{Collapse with growing asphericity\label{s3}}

A counterpart to the solution~(\ref{decx}),(\ref{decr}) comes from~\cite{2}: using the TrT, one finds:
\bea \label{grx}
x_1(t,\xi_1) = (t_0-t)^{2/3}\xi_1 + (t_0-t)^{-1/3} F(\xi_1),\qquad x_{2,3}(t,\xi_{2,3})=(t_0-t)^{2/3}\xi_{2,3}\; ;\\
\label{grr}
\rho(t,\xi_1) = \frac{1}{6\pi G (t_0-t)^2} \,\frac{1}{1 + (t_0-t)^{-1}F^{'}(\xi_1)}\; .\qquad\qquad\qquad 
\eea
 The deviation from the spherical symmetry here increases with the time (growing mode); initially  ($t=-\infty$) the flow is entirely spherical. The density is non--singular until $t=t_0$ if and only if $F^{'}(\xi_1) \geq0$. Unlike the previous two cases~(\ref{isotr}) and~(\ref{decx}),  near the collapse it is inversely proportional to the first, instead of the second, power of $(t_0-t)$ and, generically, remains non-uniform, $\rho\sim \left[6\pi G (t_0-t)F^{'}(\xi_1)\right]^{-1}$; instead of a point, the dust collapses to the $x_1$--axis.

If the last inequality is violated, and the minimum of $F^{'}(\xi_1)$ is at a single point,
\be\label{Fprneg}
\min_{\xi_1} F^{'}(\xi_1) = F^{'}(\xi_1^*)=-t_*<0\; ,
\ee
then a {\it local} collapse occurs at $t=t_0-t_*$ before the global one at $t=t_0$: the density becomes infinite at the plane $\xi_1=\xi_1^*$, or $x_1= t_*^{2/3}\xi_1^* + t_*^{-1/3} F(\xi_1^*)$. Note that $\rho(t,\xi_1^*)$ is maximum before the collapse; at the collapse, $t=t_0-t_*$, the density at all other points $\xi_1\not=\xi_1^*$ remains finite and non-zero.  

If the same minimum~(\ref{Fprneg})  is achieved at a number of discrete $\xi_1$--points, then the `early' collapse happens at all the corresponding planes. If  Eq.~(\ref{Fprneg}) holds for some $\xi_1$-- interval, 
\be\label{Fconst}
 F(\xi_1)=-t_*\xi_1, \qquad \alpha<\xi_1<\beta\; ,
\ee
then $x_1(t,\xi_1) = (t_0-t)^{-1/3}(t_0-t_*-t)\xi_1,\;x_1(t_0-t^*,\xi_1)=0$ for $\alpha<\xi_1<\beta$, and this whole $\xi_1$--layer collapses to a single plane $x_1=0$. Ultimately, when Eq.~(\ref{Fprneg}) is valid for {\it all} $\xi_1$, i.e., $F(\xi_1)\equiv-t_*\xi_1$, the early collapse becomes {\it global}: the density  $\rho(t,\xi_1) = \left[{6\pi G (t_0-t)(t_0-t_*-t)}\right]^{-1}$ is uniform, so the whole space collapses. Still, the flow remains anisotropic and the dust goes to the plane $x_1=0$, and not to a point.

\section{General solution\label{s4}}

In paper~\cite{2}, a solution containing both modes was also found; its time reversal is:
\bea \label{genx}
x_1(t,\xi_1) = (t_0-t)^{2/3}\xi_1  + (t_0-t)^{4/3} f(\xi_1)+ (t_0-t)^{-1/3} F(\xi_1)\; ;\\
\label{genr}
\rho(t,\xi_1) = \frac{1}{6\pi G (t_0-t)^2} \,\frac{1}{1 + (t_0-t)^{2/3}f^{'}(\xi_1)+ (t_0-t)^{-1}F^{'}(\xi_1)}\; 
\eea
 ($x_{2,3}$ are as before). It is general in a sense that the two free functions $f(\xi_1)$ and $F(\xi_1)$ allow one to meet any
initial conditions for both the $x_1$-coordinate and $x_1$-velocity. The solution~(\ref{genx}),(\ref{genr}) is valid at all times for 
$f^{'}(\xi_1) \geq0,\;F^{'}(\xi_1) \geq0$; other cases are analyzed as above. Of course, the decaying mode plays no role near the collapse, which is thus either an early $(t<t_0)$ planar, or a global axial one at $t=t_0$.

Note that solutions describing the expansion with an anisotropic flow and uniform density $\rho (t) = \left[6\pi G(t-t_0)\right]^{-2}$ were studied in paper~\cite{4}. Their time reversals can add to the set of non--spherical collapse solutions obtained here.

\section{Anisotropic collapse of a finite mass\label{s4a}}

All the solutions of secs.~\ref{s2} - \ref{s4} apply to circular cylinders of a finite mass with the symmetry axis along $x_1$, of an arbitrary initial radius and height, whose density is uniform in both the radial and azimuthal directions. Such a cylinder is specified by the following bounded set of Lagrange variables: 
\[
\alpha<\xi_1<\beta,\quad\sqrt{\xi_2^2+\xi_3^2}<\gamma;\qquad -\infty<\alpha<\beta<\infty,\quad 0<\gamma<\infty\; .
\]
 The decaying--mode solution of sec.~\ref{s2} describes the collapse of a cylinder to a point. Two  other solutions, of secs.~\ref{s3} and \ref{s4}, containing the growing mode,  end up with an early ($t<t_0$) collapse  to  a disk (or a number of disks) perpendicular to the $x_1$ axis, when $F^{'}(\xi_1)<0$ for some values of $\xi_1$ in the above range. Otherwise $F^{'}(\xi_1)>0$ for all $\xi_1$, and the cylinder collapses at $t=t_0$ to a part of the $x_1$--axis. Depending on the behavior of $F(\xi_1)$, it can be either the whole axis, or semi--axis, or a finite segment moving to infinity when $t\to t_0-0$. So the outcome of the collapse process is determined by the initial conditions in $x_1$--direction, if other initial conditions provide for the axial symmetry. Namely, at an initial moment of time $t=t_i$ the velocities should be  consistent with the coordinates, $v_{2,3}=-(2/3)[x_{2,3}/(t_0-t_i)]$, which relations hold then on all the way to the collapse.

\section{Collapse on the dark energy background\label{s5}}

The original Friedmann cosmological solution includes the
cosmological constant $\Lambda$, which may be zero or non-zero.  The time-reversed isotropic flow with $\Lambda>0$ is:
\be
\vec x(t,\vec\xi) = a(t)\vec\xi,  \quad \rho (t) \propto [a(t)]^{-3}; \qquad a(t)\propto\sinh^{2/3}\left[(3/2)(\sqrt{3\Lambda})(t_0 - t) \right]\; .
\ee
 The corresponding non-spherical solution of paper~\cite{3} after the TrT becomes:
\be\label{Lam}  
x_1 (t,\xi_1)= a(t)\xi_1 + \phi(t) f(\xi_1) + \Phi(t)F(\xi_1),\qquad x_{2,3}(t,\xi_{2,3})=a(t)\xi_{2,3}\; .
\ee
Functions  $\phi(t)$ and $\Phi(t)$ giving the time dependence of the decaying and growing mode are found in~\cite{3}. Solution~(\ref{Lam}) is general in the same
sense as in the case~(\ref{genx}).

\vfill\eject

\end{document}